\documentclass[pra,notitlepage,twocolumn,superscriptaddress,showpacs]{revtex4-1}

\usepackage{amsmath}
\usepackage{amsfonts}
\usepackage{graphicx}
\usepackage{color}
\usepackage{quantum}
\usepackage{epstopdf}
\usepackage{hyperref}
\usepackage[top=1.8cm, bottom=2.1cm , left=1.8cm, right=1.8cm]{geometry}

\newcommand{\mc}[1]{\mathcal{#1}}
\newcommand{\mb}[1]{\mathbf{#1}}
\newcommand{\nf}{N^*_\mc{F}}
\newcommand{\nd}{N^*_D}
\newcommand{\ghz}{\text{GHZ}}

\begin{document}
\title{Quantum macroscopicity versus distillation of macroscopic superpositions}

\author{Benjamin Yadin} \affiliation{Atomic and Laser Physics, Clarendon Laboratory, University of Oxford, Parks Road, Oxford, OX1 3PU, UK}
\author{Vlatko Vedral} \affiliation{Atomic and Laser Physics, Clarendon Laboratory, University of Oxford, Parks Road, Oxford, OX1 3PU, UK} \affiliation{Centre for Quantum Technologies, National University of Singapore, Singapore 117543}
\date{\today}
\begin{abstract}
We suggest a way to quantify a type of macroscopic entanglement via distillation of Greenberger-Horne-Zeilinger states by local operations and classical communication. We analyze how this relates to an existing measure of quantum macroscopicity based on the quantum Fisher information in several examples. Both cluster states and Kitaev surface code states are found to not be macroscopically quantum but can be distilled into macroscopic superpositions. We look at these distillation protocols in more detail and ask whether they are robust to perturbations. One key result is that one-dimensional cluster states are not distilled robustly but higher-dimensional cluster states are.
\end{abstract}


\maketitle

\section{Introduction}

Despite the overwhelming successes of quantum mechanics, one of its greatest remaining problems is to explain why it appears to break down at the macroscopic scale. In particular, macroscopic objects are never seen in quantum superpositions. The well-known thought experiment of Schr\"odinger's cat highlights the absurdity of a cat existing in a superposition of alive and dead states, yet in principle this is possible within quantum theory. It is therefore important to attempt to create macroscopic quantum states in experiments, in order to probe the boundary between quantum and classical mechanics -- to decide if a fundamental size limit exists, or if the challenge is purely a matter of isolating a system from its noisy environment.

Some recent experiments have sought `cat states' in photonic systems or similar macroscopic superpositions in superconducting circuits, molecular interferometers and mechanical resonators \cite{gao2010experimental,vanderwal2000quantum,nairz2003quantum,farrow2015classification,arndt2014testing}. Due to this great variety, one needs a general measure of `quantum macroscopicity' to compare experiments in which qualitatively different states are produced. Such a measure may also help us better understand the transition to macroscopic classical behavior.

There is no single measure generally agreed to quantify macroscopicity; typically, proposed measures are motivated along the lines of `working definition 1' given by Fr\"owis and D\"ur \cite{frowis2012measures}: a quantum state is macroscopic if it is able to display nonclassical behavior at a large scale that is not simply an accumulation of microscopic quantum effects. The need to rule out accumulated phenomena was originally appreciated by Leggett \cite{leggett1980macroscopic,leggett2002testing}. These include, for example, bulk properties of condensed matter systems that are explained only by quantum physics, yet which are built up from effects extending over the atomic scale. In other words, one expects that a macroscopic quantum state necessarily has many-body or long-range quantum correlations.

An appropriate measure should then describe the largest scale to which quantum effects extend in a given state -- this is often referred to as an `effective size', denoted here by $N^*$. We will not impose any cut-off above which a value of $N^*$ counts as macroscopic, but instead consider families of states parameterized by some obvious size quantity $N$ (e.g. the number of qubits). Then the relevant property of the family is the scaling of $N^*$ with $N$. The case $N^* = O(N)$ is `maximally macroscopic' \footnote{We use the convention $f(N)=O(g(N)) \Leftrightarrow \lim_{N \to \infty} f(N)/g(N) \in (0,\infty)$}.

In this work, we explore the consequences of viewing macroscopicity as a statement about quantum correlations. Thus we propose an effective size based on distilling macroscopic superpositions, as a way of quantifying a kind of macroscopic entanglement. Statements about entanglement are easiest for finite-dimensional systems, so our work is currently restricted to these (we focus on systems of qubits here), although characterizations of macroscopicity exist for continuous-variable systems \cite{lee2011quantification,sekatski2014size,frowis2015linking}. We compare this quantity against an existing widely-studied measure of macroscopicity based on the quantum Fisher information.

In Section \ref{sec:fisher}, we first introduce the quantum Fisher information measure of macroscopicity, then propose a measure of pure state macroscopic entanglement via distillation of Greenberger-Horne-Zeilinger (GHZ) states \cite{greenberger1989going} in Section \ref{sec:entanglement_measure}, and investigate how these relate in specific examples. We find that cluster states and Kitaev surface code ground states have macroscopic entanglement, but this is not detected by the Fisher information measure. In Section \ref{sec:robustness}, we ask whether these distillation protocols are sensitive to imperfections, finding answers via mappings onto statistical spin models. We conclude in Section \ref{sec:conclusions}.

\section{Fisher information measure of macroscopicity} \label{sec:fisher}
Given a state $\rho$ and an observable $A$, the quantum Fisher information can be defined by
\begin{equation}
	\mc{F}(\rho,A) = 2 \sum_{a,b} \frac{(p_a-p_b)^2}{p_a+p_b} \abs{\braXket{\psi_a}{A}{\psi_b}}^2,
\end{equation}
where $p_a$ and $\ket{\psi_a}$ are the eigenvalues and eigenstates of $\rho$. We consider the class $\mc{A}$ of observables which can be written as $A=\sum_{i=1}^N A_i$ over local $A_i$, each acting nontrivially on a single qubit $i$ and with fixed norm $\|A_i\|=1$. The effective size proposed by Fr\"owis and D\"ur \cite{frowis2012measures} is
\begin{equation}
	\nf(\rho) := \max_{A \in \mc{A}} \frac{\mc{F}(\rho,A)}{4N},
	\label{eqn:nf}
\end{equation}
and lies in the range $[1,N]$. ($\mc{A}$ may be extended to `$k$-local' $A$ with $A_i$ acting on groups of $k$ qubits, with $k$ bounded independent of $N$, in which case the denominator of equation (\ref{eqn:nf}) contains the number of groups instead of $N$.)

Observables in $\mc{A}$ are supposed to model the kinds of quantities that are easily measured at the macroscopic scale with coarse-grained, noisy classical detectors. For pure states, $\frac{1}{4}\mc{F}$ equals the variance, and $\nf$ is seen to quantify the largest quantum fluctuations of any macroscopic observable -- originally identified in \cite{shimizu2002stability,shimizu2005detection}. It has been shown \cite{frowis2012measures} that $\nf$ is more inclusive than a variety of measures \cite{korsbakken2007measurement,marquardt2008measuring,bjork2004size} looking at `macroscopic superpositions' of two states: maximal macroscopicity according to any of these measures implies $\nf = O(N)$.

In general, $\nf$ describes the usefulness of a state for quantum metrology. Consider a family of states $\rho_\theta = e^{-i\theta A}\rho e^{i\theta A}$ encoding a parameter $\theta \in \mathbb{R}$. From $n$ independent copies of $\rho$, the quantum Cram\'er-Rao bound sets a lower limit on the uncertainty with which $\theta$ can be estimated: $\delta \theta \geq 1/\sqrt{n\mc{F}(\rho,A)}$ \cite{braunstein1994statistical}. A macroscopic quantum state with $\nf = O(N)$ makes $\delta \theta \propto 1/N$ possible, a qualitative improvement over the classical $\delta \theta \propto 1/\sqrt{N}$.

The authors have recently provided a motivation for this measure as a quantifier of macroscopic coherence in a precise sense \cite{yadin2015general}. (That work also notes some similarities between this measure and other approaches motivated from very different starting points \cite{lee2011quantification,nimmrichter2013macroscopicity}.)

Furthermore, it has been demonstrated that large $\nf$ is a witness of macroscopic entanglement in the following ways. For pure `$k$-producible' states in which blocks of up to $k$ sites may be entangled, $\frac{1}{4}\mc{F}(\ketbra{\psi}{\psi},A) \leq kN$ for 1-local $A$ (similar bounds exist for mixtures) \cite{toth2012multipartite,hyllus2012fisher}. Also, $\nf=O(N)$ for a pure state implies that macroscopically many (i.e.\ $O(N^2)$) pairs of sites have a nonvanishing $O(1)$ amount of localizable entanglement \cite{morimae2010superposition,popp2005localizable}.

However, the converse is false: there are highly-entangled states with small $\nf$. To put this precisely, we will introduce a measure aiming to quantify macroscopic entanglement.

\section{Measure of macroscopic entanglement} \label{sec:entanglement_measure}
For a pure quantum state, one can quantify the amount of bipartite entanglement by counting the number of maximally entangled states that can be distilled by local operations and classical communication (LOCC) from many copies of the given state \cite{bennett1996concentrating}.

In the multipartite case there is no unique maximally entangled state \cite{dur2000three}. To give a reasonable notion of macroscopic entanglement, we suggest to use the GHZ states $\ket{\ghz_n} := (\ket{0}^{\ox n} + \ket{1}^{\ox n})/\sqrt{2}$ as the target for distillation. $\ket{\ghz_n}$ is often described as the typical qubit model of a macroscopic superposition, and is the maximally macroscopic state of $n$ qubits in the sense that $\nf=n$. Furthermore, it is easy to motivate assigning an effective size of $n$ to such a state.

To define our measure, take a pure state $\ket{\psi}$ of $N$ qubits, and consider acting on $\ket{\psi}$ with stochastic LOCC (SLOCC), described by measurement operators $\{M_a\}$ corresponding to the outcomes $\sqrt{p_a} \ket{\phi_a} := M_a \ket{\psi}$ with probabilities $p_a = \braXket{\psi}{M_a^\dagger M_a}{\psi}$. We denote this kind of transformation by $\ket{\psi} \to \{\ket{\phi_a},p_a\}$. Now restrict these operations to the set $D_\psi$ such that every outcome is of the form $\ket{\phi_a} = \ket{\ghz_{S_a}}\ket{0}^{N-\abs{S_a}}$ where $S_a \subseteq \{1,2,\dots,N\}$ is some subset of $N$ qubits of cardinality $\abs{S_a}$. We associate with each $\ket{\phi_a}$ a size $n_a = \abs{S_a}$ unless a `trivial' GHZ state of size 1 is obtained, in which case $n_a=0$. Our measure is

\begin{equation}
	\nd(\ket{\psi}) := \max_{\{M_a\} \in D_\psi} \sum_a p_a n_a.
	\label{eqn:nd}
\end{equation}

This is supposed to describe the size of GHZ-type entanglement present in the state $\ket{\psi}$. We have restricted each final state to a single GHZ, rather than a general product $\bigotimes_i \ket{\ghz_{n_i}}$, since we are only interested in the largest GHZ; the remaining parts could be converted deterministically into product states. This prescription, instead of summing the sizes, rules out the `accumulated' phenomena mentioned earlier. Thus a state like $(\ket{00}+\ket{11})^{\ox N}$ has $\nd = O(1)$ instead of $O(N)$. In general, $\nd(\bigotimes_i \ket{\psi_i}) = \max_i \nd(\ket{\psi_i})$.

It is simple to show that $\nd$ is an entanglement monotone -- it cannot increase on average under SLOCC. Suppose $\ket{\psi} \to \{\ket{\chi_\mu}, p_\mu \}$ by SLOCC. Then for each $\mu$ there exists an optimal ensemble $\{\ket{\phi_{\mu,a}}, p_{\mu,a} \}$ distilled from $\ket{\chi_\mu}$ such that
\begin{equation}
	\nd(\ket{\chi_\mu}) = \sum_a p_{\mu,a} n_{\mu,a},
\end{equation}
where $\ket{\phi_{\mu,a}}$ contains a GHZ state of size $n_{\mu,a}$. By composing the two SLOCC protocols, it follows that $\ket{\psi} \to \{\ket{\phi_{\mu,a}}, p_\mu p_{\mu,a} \}$ is a valid distillation operation in $D_\psi$, so
\begin{align}
	\nd(\ket{\psi}) & \geq \sum_{\mu,a} p_\mu p_{\mu,a} n_{\mu,a} \nonumber \\
	& = \sum_\mu p_\mu \nd(\ket{\chi_\mu}),
\end{align}
which proves the monotonicity.

The optimization involved in determining $\nd$ will generally be intractable -- the best we can do is find bounds. A lower bound must come from an explicit construction of distillation operations, and this will be difficult except for some particular cases. By extending a method in \cite{dur2002effective}, we present a simple upper bound for states that are symmetric under exchange of any two sites (see Appendix \ref{app:nd_upper_bound}). This comes from the geometric entanglement \cite{wei2003geometric} of a single site with the rest: $E_G(\ket{\psi}) = 1 - \lambda_\text{max}(\ket{\psi})$, where $\lambda_\text{max}$ is the largest eigenvalue of $\rho_1 := \tr_{2,3,\dots,N} \ketbra{\psi}{\psi}$. Using the monotonicity of $E_G$ under SLOCC, we find
\begin{equation}
	\nd(\ket{\psi}) \leq 2N (1 - \lambda_\text{max}(\ket{\psi}) ).
	\label{eqn:upper_bound}
\end{equation}

\subsection{Generalized GHZ states}
Generalized GHZ states \cite{dur2002effective} were our initial motivation for considering GHZ distillation. These depend on $N$ and a parameter $\epsilon \in \mathbb{R}$, and were suggested to be a reasonable description of the macroscopic current superpositions in superconducting qubits. They can be written as $\ket{\psi_N^\epsilon} \propto \ket{\epsilon}^{\ox N} + \ket{-\epsilon}^{\ox N}$, where $\ket{\pm \epsilon} := \cos(\epsilon/2) \ket{0} \pm \sin(\epsilon/2) \ket{1}$. For $\epsilon=\frac{\pi}{2}$ we recover $\ket{\ghz_N}$, while $\epsilon=0$ gives $\ket{0}^{\ox N}$. Thus we expect $\epsilon$ to vary the macroscopicity smoothly between the minimal and maximal values. Indeed, for $\epsilon \ll 1 \ll N\epsilon^2$, $\nf \approx N \epsilon^2$ \cite{frowis2012measures}.

The distillation protocol constructed in \cite{dur2002effective} produces an average distilled size of approximately $N\epsilon^2/2$ in the above limit. Moreover, $\lambda_\text{max} \approx \cos^2(\epsilon/2)$ giving $\nd \leq N\epsilon^2/2$ -- so the protocol is exactly optimal \footnote{This is an improvement over the bound $N\epsilon^2 \log_2(1/\epsilon)/2$ in \cite{dur2002effective}.}, and $\nd(\ket{\psi_N^\epsilon}) \approx N\epsilon^2/2$. Here we have $\nd \approx \nf / 2$.

\subsection{Cluster states}
It has already been noted for cluster states that $\nf=O(1)$, but it is nevertheless possible to deterministically distill from them GHZ states of size $O(N)$ \cite{frowis2012measures,frowis2015linking}. Recall that a $d$-dimensional cluster state is defined for qubits associated with the vertices of (a subset of) a $d$-dimensional square lattice. It can be constructed from $(\ket{0}+\ket{1})^{\ox N}$ by applying a controlled-$\sigma^z$ gate between each neighboring pair \cite{briegel2001persistent}.

To see the scaling of $\nf$, note that the variance quantifies the total amount of two-point correlations \cite{morimae2010superposition}. In cluster states, it can be shown that two regions are uncorrelated unless they share a boundary. Since each region of bounded $O(1)$ size has a bounded number of neighbors, it follows that $\nf = O(1)$.

\begin{figure}[h]
	\centering
	\includegraphics[scale=1]{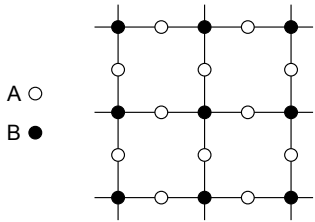}
	\caption{The graph used to define the cluster state $\ket{C_N}$. Measurements are performed on the $A$ sites in order to distill the $B$ sites into a GHZ state.}
	\label{fig:cluster}
\end{figure}

For simplicity, we focus on the family of $N$-qubit cluster states $\ket{C_N}$ defined by graphs of the type shown in fig.\ \ref{fig:cluster} -- these are two-dimensional cluster states with a fraction of sites removed. A measurement of each site of set $A$ in the $\sigma^x$ eigenbasis projects $B$ into $\ket{\ghz_{N_B}}$ up to local $\sigma^x$ gates depending on the outcomes, where $N_B$ is the number of $B$ sites. Classical communication can remove these errors, making the final state deterministically $\ket{\ghz_{N_B}}$. Therefore $\nd(\ket{C_N}) \geq N_B = O(N)$. It is simple to see that this generalizes to cluster states of any dimension.

\subsection{Kitaev surface code states}
Kitaev's surface code model is the simplest lattice system displaying topological order, of great interest for condensed matter physics and topological quantum computation \cite{kitaev2003fault,wen2004quantum,bombin2013introduction}. Its ground states are sensitive to \emph{global} topological properties of their lattice. Since they could exist on a macroscopic lattice, this makes them interesting candidates for macroscopicity.
\begin{figure}[h]
	\centering
	\includegraphics[scale=1]{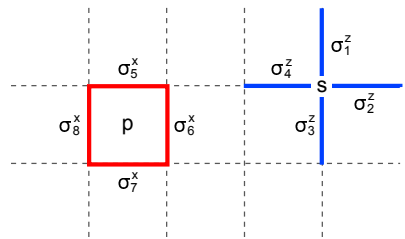}
	\caption{A star $s$ and plaquette $p$ in Kitaev's model, and their corresponding operators: $z_s = \sigma^z_1 \sigma^z_2 \sigma^z_3 \sigma^z_4$ and $x_p = \sigma^x_5 \sigma^x_6 \sigma^x_7 \sigma^x_8$.}
	\label{fig:kitaev}
\end{figure}

The two-dimensional version is defined by a square lattice where each edge represents a qubit, with Hamiltonian $H = - \sum_s z_s - \sum_p x_p$. Here, $z_s$ is the product of $\sigma^z$ operators over a `star' and $x_p$ is the product of $\sigma^x$ operators over a `plaquette' -- see fig.\ \ref{fig:kitaev} -- and the sums are over all stars and plaquettes in the lattice. There may be a number of degenerate ground states, depending on the topology of the lattice; however, these have equivalent entanglement properties \cite{orus2014geometric} so we concentrate on the simplest ground state.

It is possible to describe the structure of this state explicitly. We define a `configuration' of the lattice to be a product state where the qubits lying on a certain set of curves (open or closed) are all in the state $\ket{1}$ while the rest are $\ket{0}$. The ground state $\ket{K_N}$ is an equal superposition of all configurations containing only closed curves which are topologically trivial (contractible to a point).

$\ket{K_N}$, like $\ket{C_N}$, has no correlations between non-neighboring regions \cite{lee2013entanglement}, so $\nf = O(1)$. As a further parallel, one can deterministically distill macroscopic GHZ states from $\ket{K_N}$ by SLOCC: choose any topologically trivial non-self-intersecting loop $B$ of $N_B=O(N)$ qubits and perform local projective measurements on the remainder of the lattice in the $z$ basis. Consider the case where the outcome of each measurement is $\ket{0}$ -- then $B$ ends up in $\ket{\ghz_{N_B}}$ containing the only two consistent closed-curve configurations. For different outcomes, the final state differs by local $\sigma^x$ operations, so we can obtain $\ket{\ghz_{N_B}}$ deterministically. Hence $\nd(\ket{K_N}) = O(N)$.

\subsection{Dicke states}
Dicke states \cite{dicke1954coherence} have recently been studied for their interesting multipartite entanglement properties \cite{frowis2012measures,frowis2012cloned,bergmann2013entanglement,toth2012multipartite,novo2013genuine}. The $N$-qubit versions $\ket{N,k}$ are defined by symmetrizing $\ket{0}^{\ox(N-k)} \ket{1}^{\ox k}$. For any observable of the form $A = \sum_i \cos(\theta)\sigma^x_i + \sin(\theta)\sigma^y_i$ we obtain the maximal variance giving $\nf = 1 + 2f(1-f)N$ where the quantity $f=k/N$ controls the macroscopicity in the same way as $\epsilon$ for $\ket{\psi_N^\epsilon}$.

Our upper bound gives $\nd(\ket{N,k}) \leq 2\min\{k,N-k\}$. We are not aware of any SLOCC protocols to distill GHZ states from $\ket{N,k}$, and so cannot provide a lower bound. We can only state that $\nd \leq 2(\nf-1)$, using $\min\{f,1-f\} \leq 2f(1-f)$. Thus $\nf < O(N) \Rightarrow \nd < O(N)$. However we do not know if $\nf = O(N) \Rightarrow \nd = O(N)$.

\section{Robustness of distillation} \label{sec:robustness}
In the above cases of cluster states and Kitaev surface code states, the distillation accomplishes something of practical value: it extracts states which are useful for quantum metrology from initial states which are not, without increasing entanglement. Equivalently, it amplifies the macroscopicity of the states according to the measure $\nf$. If this is to be regarded as something with practical significance, then the final states must be robust with respect to imperfections in the protocol. Indeed, one might doubt the physical meaning of the distillation if it is not robust in this sense.

To be specific, we neglect environmental noise since the GHZ states produced are maximally sensitive to decoherence \cite{dur2002effective}, so any experiment taking advantage of them must have tolerably low noise. Instead, we suppose that the measurement device operating on the individual qubits may couple to them imprecisely. Therefore we say that:

\emph{The distillation of GHZ states is robust if the average Fisher information of the final state retains the same scaling with $N$ for any small perturbation to the measurement operators.}

The definition of the perturbation will be clear from the following examples.

\subsection{Cluster states}
For cluster states, we perturb the projective measurements to generalized measurement operators $\{E,\bar{E}\}$ satisfying $E^\dagger E + \bar{E}^\dagger \bar{E} = I$. Via the action of controlled-$\sigma^z$ gates, the initial state can be written as $\ket{C_N} \propto \sum_\mathbf{b} \ket{\mb{a}(\mb{b})_A^x} \ket{\ghz(\mb{b})_B^z}$ where $A$ and $B$ are expressed in the $x-$ and $z-$bases respectively, $\mb{a} \in \{0,1\}^{N_A}$, $\mb{b} \in \{0,1\}^{N_B}$ and $\ket{\ghz(\mb{b})}:= (\ket{\mb{b}}+\ket{\mb{\overline{b}}}) / \sqrt{2}$ with $\overline{b_i}:=1-b_i$. Each value of $\mb{a}$ is determined by $\mb{b}$, and without loss of generality we fix $b_1=0$ -- see Appendix \ref{app:cluster} for details.

In the unperturbed case, projective measurements on $A$ give full GHZ states on $B$ with equal probability for each outcome. From this symmetry it is clear that we need examine only a single branch, say the outcome $E^{\ox N_A}$. The final state is $\ket{\psi} \propto \sum_\mb{b}(E^{\ox N_A}\ket{\mb{a}(\mb{b})_A^x})\ket{\ghz(\mb{b})_B^z}$.

The most general form (up to normalization) for $E^\dagger E$ in the $x$-basis is
\begin{equation}
	E^\dagger E = 
\begin{pmatrix}
	1 & \delta^* \\
	\delta & \epsilon
	\end{pmatrix},\,
	{\delta \in \mathbb{C}},\, {\epsilon \in \mathbb{R}}.
\end{equation}
For the observable $Z_B = \sum_{i\in B} \sigma_i^z$, $\mc{F}$ depends on $\epsilon$ but not $\delta$:
\begin{equation}
	\frac{1}{4} \mc{F}(\ketbra{\psi}{\psi},Z_B) = \frac{1}{\mc{Z}} \sum_\mb{b} \epsilon^{\abs{\mb{a}(\mb{b})}} (N_B - 2\abs{\mb{b}})^2,
\end{equation}
where $\mc{Z} = \sum_\mb{b} \epsilon^{\abs{\mb{a}(\mb{b})}}$ and $\abs{\mb{a}} := \sum_i a_i$. We interpret this by mapping the problem onto a two-dimensional square-lattice ferromagnetic classical Ising model. Each $\mb{b}$ maps onto a spin configuration with individual magnetic moments of $\pm 1$ and a total magnetization $M = N_B - 2\abs{\mb{b}}$, and $\mb{a}(\mb{b})$ is the corresponding bond configuration. There is an energy cost of $2J$ for each $a_i=1$, so we associate $\epsilon$ with $e^{-2\beta J}$ and $\mc{Z}$ with the partition function. Therefore $\frac{1}{4}\mc{F}(\ketbra{\psi}{\psi},Z_B)= \expect{M^2}$.

This model has a low-temperature ferromagnetic phase; from a bound by Griffiths \cite{griffiths1964peierls} we obtain $\frac{1}{4}\mc{F}(\ketbra{\psi}{\psi},Z_B) \geq N_B^2(1-O(\epsilon^4))$ -- so the distillation is robust.

This fails for one-dimensional cluster states, because the Ising model has no $T>0$ phase transition in one dimension \cite{yeomans1992statistical}. So distillation from one-dimensional cluster states is not robust.

\subsection{Kitaev surface code states}
$\ket{K_N}$ works similarly, and the closed-loop restriction lets us interpret the configurations as domain walls of an Ising model with a spin on each plaquette. The robustness of distillation again relies on $T>0$ ferromagnetic order. However, the relevant statistical model is more complex and depends on $\delta$. As discussed in Appendix \ref{app:kitaev}, the disorder associated with $\delta$ maps onto probabilistically `turning off' the bonds, giving a dilute Ising model. This has a ferromagnetic phase \cite{thorpe1976thermodynamics}, letting us conclude $\expect{Z_B^2} = O(N_B^2)$ for sufficiently small $\delta$ and $\epsilon$.

An important caveat is that this only works when the loop $B$ divides the lattice into two-dimensional regions -- for the same reason as the failure of one-dimensional cluster states. Therefore it seems we can only conclude robustness when $N_B = O(\sqrt{N})$. The distilled states, while robust, have $\mc{F}=O(N)$ and do not offer a metrological advantage over the initial state.

\section{Conclusions} \label{sec:conclusions}
In summary, we have proposed a notion of macroscopic entanglement which measures the size of GHZ states that can be extracted by SLOCC, and compared this with an existing measure of macroscopicity. We find that cluster states and Kitaev model ground states are macroscopically entangled by our definition but are not macroscopically quantum by the Fisher information. However, we lack statements examining the converse: whether $\nf = O(N) \Rightarrow \nd = O(N)$. Extensions to our measure beyond qubits are also important -- for finite-dimensional systems, one could distill $\ket{\ghz_n^k}=\frac{1}{\sqrt{k}}\sum_{i=0}^{k-1} \ket{i}^{\ox n}$. We suggest ascribing the same size to this as to $\ket{\ghz_n^2}$, as $\ket{\ghz_n^2}^{\ox m}$ looks like $\ket{\ghz_n^{2^m}}$ at the $m$-qubit level.

The distillation can also be interpreted as increasing the advantage of these states in high-precision parameter estimation. A criterion requiring the distillation to be of practical utility, by being robust against errors, tells us that two-dimensional, but not one-dimensional, cluster states are useful. This dependence on dimensionality also appears in the statement that two-dimensional cluster states are universal for measurement-based quantum computation, while the one-dimensional versions are not \cite{nielsen2006cluster}. Kitaev model ground states are robust, but this can be proved only up to a non-macroscopic $O(\sqrt{N})$ distilled size. It is noteworthy that our analysis for these states is very close to the proof that surface codes can be error corrected \cite{dennis2002topological}. Thus we speculate that our results might be related to the ability of these states to encode quantum information and do so robustly.

As suggested by others \cite{frowis2015linking}, it would be helpful to develop a resource theory for macroscopicity, requiring an understanding of the operations unable to increase the effective size. The SLOCC distillation operations used here are not included in the set of `free' operations for macroscopic coherence defined in \cite{yadin2015general}, under which the Fisher information cannot increase. The crucial point is that the final $\sigma^x$ gates, conditioned on measurement outcomes, are not free in that framework and instead entail manipulation of coherence between eigenstates of $Z_B$. This is where the present notion of macroscopic entanglement departs from macroscopic coherence.

\begin{acknowledgments}
The authors acknowledge funding from the National Research Foundation (Singapore), the Ministry of Education (Singapore), the EPSRC (UK), the Templeton Foundation, the Leverhulme Trust, the Oxford Martin School and Wolfson College, University of Oxford, and thank Wonmin Son for helpful discussions.
\end{acknowledgments}

\appendix

\section{Upper bound on $\nd$ for symmetric states} \label{app:nd_upper_bound}
First note that for symmetric states, we may always construct a symmetric optimal distillation protocol: if an asymmetric optimal protocol is found, one just needs to probabilistically perform all its permutations with equal weighting. As a result, the probability $p_a$ for each $\ket{\ghz_{S_a}}$ is a function of $n_a$ only. We denote by $q_n$ the total probability of obtaining any GHZ state of size $n$ -- it is clear that $q_n = {N \choose n} p_a$ for any $a$ such that $\abs{S_a} = n$.

For a distilled state $\ket{\phi_a} = \ket{\ghz_{S_a}} \ket{0}^{\ox N-\abs{S_a}}$ the geometric entanglement $E_G$ of a site $i$ with the rest of the system is $\frac{1}{2}$ if $i \in S_a$ and 0 otherwise. The number of subsets of size $n$ containing $i$ is ${N-1 \choose n-1}$.

The average $E_G$ for the final ensemble is therefore
\begin{align}
	\sum_a p_a E_G(\ket{\phi_a}) & = \sum_{n=2}^N q_n \frac{{N-1 \choose n-1}}{{N \choose n}} \times \frac{1}{2} \nonumber \\
	&= \frac{1}{2}\sum_{n=2}^N q_n \frac{n}{N} \nonumber \\
	&= \frac{1}{2N} \nd(\ket{\psi}).
\end{align}
By the monotonicity of $E_G$ under SLOCC, we have
\begin{equation}
	\nd(\ket{\psi}) \leq 2N E_G(\ket{\psi}).
\end{equation}

\section{Distillation from cluster states $\ket{C_N}$} \label{app:cluster}
To construct $\ket{C_N}$, we define $\{\ket{0^x},\ket{1^x}\} := \{\ket{+},\ket{-}\}$ and write the state $\ket{+}^{\ox N}$ before applying the $U_{CZ}$ gates as
\begin{equation}
	\bigotimes_{i \in A} \ket{0_i^x} \ox \bigotimes_{j \in B} (\ket{0_j^z} + \ket{1_j^z}) = \sum_\mb{b} \ket{\mb{0}_A^x} \ox \ket{\mb{b}_B^z}.
\end{equation}
The action of $U_{CZ}$ on two qubits $i,\,j$ is determined by
\begin{equation}
	U_{CZ} \ket{a_i^x}\ket{b_j^z} = \ket{(a_i \oplus b_j)^x} \ket{b_j^z},
\end{equation}
where $\oplus$ denotes addition modulo 2. Therefore we have that
\begin{equation}
	\ket{C_N} \propto \sum_\mb{b} \ket{\mb{a}(\mb{b})_A^x} \ket{\mb{b}_B^z},
\end{equation}
where each $\mb{a}(\mb{b})$ is determined by the following rule: $a_i=0$ when the neighboring $b_i$ are equal, and $a_i=1$ when they are different. One can see that $\mb{a}(\mb{b^{(1)}}) = \mb{a}(\mb{b^{(2)}})$ if and only if $\mb{b^{(2)}} = \mb{b^{(1)}}$ or $\mb{\overline{b^{(1)}}}$. This leads to
\begin{equation}
	\ket{C_N} \propto \sum_\mb{b} \ket{\mb{a}(\mb{b})_A^x} \ket{\ghz(\mb{b})_B^z},
\end{equation}
where we fix $b_1=0$ without loss of generality -- any single site in $B$ could be fixed. It is clear from this expression that a measurement of every $A$ site in the $x$-basis will project $B$ into a GHZ state; there are $2^{N_B-1}$ different outcomes.

With imperfect measurements, as discussed in the main text, we only need to consider a single branch with the outcome $E$ for each measurement, so we use the final state
\begin{equation}
	\ket{\psi} = \frac{1}{\sqrt{\mc{Z}}} \sum_\mb{b} (E^{\ox N_A} \ket{\mb{a}(\mb{b})_A^x}) \ket{\ghz(\mb{b})_B^z}.
\end{equation}
To calculate the Fisher information in the variable $Z_B = \sum_{i \in B} \sigma_i^z$, we just need the variance $\braXket{\psi}{Z_B^2}{\psi} - \braXket{\psi}{Z_B}{\psi}^2$. Now it is easy to see from symmetry that $\braXket{\psi}{Z_B}{\psi} = 0$, while
\begin{align}
	\sum_i \sigma_i^z (\ket{\mb{b}^z} \pm \ket{\mb{\overline{b}}^z}) &= \sum_i (-1)^{b_i} (\ket{\mb{b}^z} \mp \ket{\mb{\overline{b}}^z}) \nonumber \\
		&= (N_B - 2\abs{\mb{b}}) (\ket{\mb{b}^z} \mp \ket{\mb{\overline{b}}^z})
\end{align}
gives $Z_B^2 \ket{\ghz(\mb{b})_B^z} = (N_B - 2\abs{\mb{b}})^2 \ket{\ghz(\mb{b})_B^z}$. Therefore
\begin{align}
	\braXket{\psi}{Z_B^2}{\psi} &= \frac{1}{\mc{Z}} \sum_{\mb{b},\mb{b'}} \braXket{\mb{a}(\mb{b})}{(E^\dagger E)^{\ox N_A}}{\mb{a'}(\mb{b'})} \nonumber \\
		& \qquad \times  \braXket{\ghz(\mb{b})_B^z}{Z_B^2}{\ghz(\mb{b'})_B^z} \nonumber \\
		&= \frac{1}{\mc{Z}} \sum_{\mb{b},\mb{b'}} \braXket{\mb{a}(\mb{b})}{(E^\dagger E)^{\ox N_A}}{\mb{a'}(\mb{b'})} \nonumber \\
		& \qquad \times \, \delta_{\mb{b},\mb{b'}} (N_B - 2\abs{\mb{b}})^2 \nonumber \\
		&= \frac{1}{\mc{Z}} \sum_\mb{b} \epsilon^{\abs{\mb{a}(\mb{b})}} (N_B - 2\abs{\mb{b}})^2,
\end{align}
given that $\braXket{0}{E^\dagger E}{0} = 1,\, \braXket{1}{E^\dagger E}{1} = \epsilon$. Similarly, by setting $\braket{\psi}{\psi}=1$ we obtain $\mc{Z} = \sum_\mb{b} \epsilon^{\abs{\mb{a}(\mb{b})}}$.

As described in the main text, $\frac{1}{4}\mc{F}(\ketbra{\psi}{\psi},Z_B)$ can be interpreted as $\expect{M^2}$ for a classical Ising model with $\epsilon = e^{-2\beta J}$. Griffiths \cite{griffiths1964peierls} establishes a bound of the form $\expect{\abs{M}} \geq N_B(1-f(T))$ where $f$ is independent of $N_B$ and $\lim_{T\to 0} f(T) = 0$. Using $0\leq \expect{(N_B - \abs{M})^2} = N_B^2 - 2N_B\expect{\abs{M}} + \expect{M^2}$, it follows that $\expect{M^2} \geq 2N_B\expect{\abs{M}} - N_B^2 \geq N_B^2 (1 - 2f(T))$. To leading order, $f(T) \approx e^{-8\beta J}$ which gives the bound

\begin{equation}
	\frac{1}{4} \mc{F}(\ketbra{\psi}{\psi},Z_B) \geq N_B^2 (1 - O(\epsilon^4)).
\end{equation}

\section{Distillation from Kitaev model ground states $\ket{K_N}$} \label{app:kitaev}
\begin{figure}[h]
	\centering
	\includegraphics[scale=1]{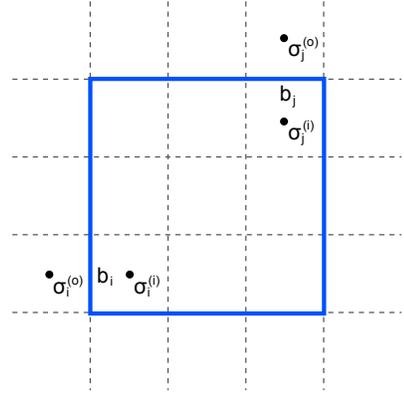}
	\caption{The highlighted loop $B$ is the set of qubits to be distilled into a GHZ state. Each segment $b_i$ has neighboring Ising spins $\sigma^{(i)}_i$ and $\sigma^{(o)}_i$.}
	\label{fig:kitaev-distillation}
\end{figure}

As above, we need to calculate $\expect{Z_B^2}$ and verify that it remains $O(N_B^2)$ for sufficiently small perturbations. Our approach will again involve a map onto a statistical model where a spin $\sigma_i$ is placed at the center of each plaquette. $B$ is chosen to be a rectangular loop of size $N_B = O(\sqrt{N})$ which cuts the remainder $A$ of the lattice into two independent two-dimensional regions. If we label each edge in $B$ by $b_i=0,1$ with neighboring `inside' and `outside' Ising spins $\sigma_i^{(i)}, \sigma_i^{(o)} = \pm 1$ (see fig.\ \ref{fig:kitaev-distillation}), then $\sigma_i^{(i)} \sigma_i^{(o)} = 1 - 2b_i$ and $\expect{Z_B^2}$ will be given by
\begin{align}
	\expect{(N_B - 2\abs{\mb{b}})^2} &= \expect{(\sum_{i \in B} \sigma_i^{(i)} \sigma_i^{(o)})^2} \nonumber \\
		&= N_B + 2 \sum_{i<j} \expect{\sigma_i^{(i)}\sigma_j^{(i)}} \expect{\sigma_i^{(o)}\sigma_j^{(o)}},
\end{align}
where the expectation value is taken with respect to the statistical model to be defined below. Hence it will follow that $\expect{Z_B^2} = O(N_B^2)$ if our model results in an ordered phase where the two-point correlation functions are bounded strictly above zero independent of $N$.

To define the required mapping, we recall that the pre-measurement state is a sum of (topologically trivial) closed-loop configurations in the $z$-basis:
\begin{equation}
	\ket{K_N} \propto \sum_{c:\,\partial c=0} \ket{c_A}\ket{c_B},
\end{equation}
where $\partial c$ is the boundary of $c$ and $c_A$ and $c_B$ are the parts of $c$ existing on $A$ and $B$ respectively. (For the remainder of this section, a quantity such as $c$ or $c_A$ will be understood to be a vector, while $c_i$ is a single component.) Since $B$ is a closed loop, every distinct $c_A$ in this sum corresponds to exactly two values of $c_B$, which are the opposites of each other. Therefore, by choosing to fix a single site of $B$, say $b_1=0$ (which we will do implicitly from now on), we can write the state as
\begin{equation}
	\ket{K_N} \propto \sum_{c:\,\partial c=0} \ket{c_A} \ket{\ghz(c_B)}.
\end{equation}
Furthermore, for each $c_B$ we can generate all the corresponding $c_A$ by choosing a particular representative $c_A(c_B)$ and adding to this all the possible closed loops $z_A$ lying strictly within $A$. (Note that adding $z_A$ to $c_A$ creates a configuration $\ket{c_A \oplus z_A}$ in which $c_A$ is deformed through the region bounded by $z_A$.) This provides an additional representation
\begin{equation}
	\ket{K_N} \propto \sum_{c_B} \left( \sum_{z_A:\,\partial z_A=0} \ket{c_A(c_B) \oplus z_A} \right) \ket{\ghz(c_B)}.
\end{equation}
Hence we can write the post-measurement state $E^{\ox N_A}\ket{K_N}$ as
\begin{align}
	\ket{\psi} &= \frac{1}{\sqrt{\mc{Z}}} \sum_{c:\,\partial c=0} E^{\ox N_A}\ket{c_A} \ket{\ghz(c_B)} \\
		&= \frac{1}{\sqrt{\mc{Z}}} \sum_{c_B}  \left( \sum_{z_A:\,\partial z_A=0} E^{\ox N_A}\ket{c_A(c_B) \oplus z_A} \right) \nonumber \\
		& \qquad \times \ket{\ghz(c_B)}
\end{align}
such that $\mc{Z}$ gives the normalization $\braket{\psi}{\psi}=1$. We can use these two forms simultaneously to determine $\mc{Z}$:
\begin{align}
	\mc{Z} &= \sum_{c:\,\partial c=0} \; \sum_{c_B'}\; \sum_{z_A:\,\partial z_A=0} \braXket{c_A'(c_B') \oplus z_A}{\mc{E}}{c_A} \nonumber \\
		&  \hspace{8em} \times \braket{\ghz(c_B')}{\ghz(c_B)} \nonumber \\
		&= \sum_{c:\,\partial c=0} \; \sum_{z_A:\,\partial z_A=0} \braXket{c_A \oplus z_A}{\mc{E}}{c_A},
\end{align}
where $\mc{E}:= (E^\dagger E)^{\ox N_A}$. In the second line, we have used the fact that $c'_B=c_B$ to choose the representative $c'_A(c'_B)$ to equal $c_A$. Similarly we find
\begin{equation}
	\expect{Z_B^2} = \frac{1}{\mc{Z}} \sum_{c:\,\partial c=0}\; \sum_{z_A:\,\partial z_A=0} \braXket{c_A \oplus z_A}{\mc{E}}{c_A} (N_B-\abs{c_B})^2.
\end{equation}
From the general form of $E^\dagger E$ described in the main text, we have \footnote{We take $\delta$ to be real without loss of generality, as the sums for $\mc{Z}$ and $\expect{Z_B^2}$ will pick out the real part in any case.}
\begin{equation}
	\braXket{c_A \oplus z_A}{\mc{E}}{c_A} = \prod_{i \in A} \delta^{z_i} \epsilon^{c_i(1-z_i)}.
\end{equation}

We shall see that this calculation can be mapped onto a statistical spin model with the following Hamiltonian:
\begin{equation}
	H = \sum_{<i,j>} J_{ij} \sigma_i \sigma_j,
\end{equation}
where the sum is over neighboring pairs of spins, and $J_{ij} = 0$ with probability $p$ and $J_{ij} = - J \; (J>0)$ with probability $1-p$. Such a model describes an Ising ferromagnet with random disorder caused by removing some of the bonds; we take this disorder to be `annealed' (as opposed to `quenched'), meaning that the $J_{ij}$ are considered dynamical variables. It is known that this model is in a ferromagnetic phase for sufficiently small $T$ and $p$ \cite{thorpe1976thermodynamics}.

Let us first rewrite the Hamiltonian in terms of the variables $f_{ij} = 0,1$ with probabilities $(1-p),p$ respectively, and $b_{ij} = \frac{1}{2}(1-\sigma_i \sigma_j)$:
\begin{equation}
	H(f,b) = -J \sum_e (1-f_e)(1-2b_e),
\end{equation}
where we replace $(ij)$ with $e$ labeling the edges of the lattice dual to the spins. Up to normalization, the probability of a configuration $(f,b)$ is
\begin{align}
	P(f,b) &\propto e^{-\beta H(f,b)} \prod_e \left(\frac{p}{1-p}\right)^{f_e} \nonumber \\
		&\propto \prod_e e^{-\beta J(f_e + 2b_e(1 - f_e))} \left(\frac{p}{1-p}\right)^{f_e} \nonumber \\
		&= \prod_e \left[\left(\frac{p}{1-p}\right) e^{-\beta J} \right]^{f_e} (e^{-2\beta J})^{b_e (1-f_e)}.
\end{align}
Therefore we can make the identification $\braXket{c_A \oplus z_A}{\mc{E}}{c_A} \to P(f,b)$ as long as we map
\begin{align}
	\delta & \to \left(\frac{p}{1-p}\right) e^{-\beta J}, \nonumber \\
	\epsilon & \to   e^{-2\beta J}, \nonumber\\
	z_i & \to  f_e, \nonumber \\
	c_i & \to  b_e.
\end{align}
The last of these is compatible with the restriction $\partial c=0$ because each bond configuration $b$ for the spins must be formed of closed loops. However, our condition that $\partial z_A=0$ means that we must similarly restrict the patterns of $f$ in the spin model; it is nevertheless clear that this only reduces the disorder and so must preserve the ferromagnetic phase. There are also no bonds lying on the curve $B$, explaining our earlier claim that $B$ cuts the remainder $A$ into two non-interacting sections.

Therefore, to sum up, each term $\braXket{c_A \oplus z_A}{\mc{E}}{c_A}$ can be interpreted as the probability of a configuration in this disordered spin model, so that $\mc{Z}$ becomes the partition function and $\expect{Z_B^2}$ is related to the two-point correlators as described earlier. The existence of a ferromagnetic phase then lets us conclude that $\expect{Z_B^2} = O(N_B^2)$ for sufficiently small $\delta$ and $\epsilon$.

\begin{figure}
	\centering
	\vspace{10pt}
	\includegraphics[scale=0.7]{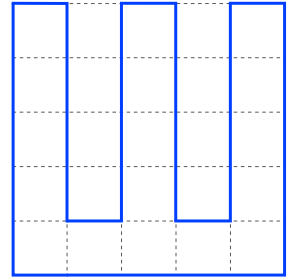}
	\caption{An example of a loop to be distilled into a size $O(N)$ GHZ state.}
	\label{fig:kitaev-1d}
\end{figure}

This argument assumed $B$ to be a rectangular loop of size $O(\sqrt{N})$, such that the distilled GHZ states have a Fisher information of only $O(N)$ -- the same as the initial state $\ket{K_N}$. If we want an improvement for metrology, we would need to choose a loop of size $O(N)$. As depicted in fig.\ \ref{fig:kitaev-1d}, it seems to us that any such loop will divide the lattice into one-dimensional rather than two-dimensional regions, in which case the above argument does not apply. By analogy with the cluster state example, we might conjecture that no $O(N)$ loop is robust to perturbations.

\bibliographystyle{h-physrev}
\bibliography{macroscopicity}
\end{document}